\documentstyle[11pt,epsfig]{article}
\author {M. Neek-Amal$^1$~  and  F. M. Peeters$^2$ \\
\small $^1$ Department of Physics, Shahid Rajaee Teacher Training
University,
Lavizan, Tehran 16788, Iran.\\
 {\small $^2$Departement Fysica, Universiteit
Antwerpen, Groenenborgerlaan 171, B-2020 Antwerpen,
 Belgium. }}
\begin{document}
\title{\bf Graphene nano ribbons subjected to axial stress}
\date{\today}
\maketitle
\begin{abstract}
Atomistic simulations are used to study the bending of rectangular
graphene nano ribbons subjected to  axial stress  both for free
boundary and supported boundary conditions. The shape of the
deformations of the buckled graphene nano ribbons, for small values
of the stress, are sine waves where the number of nodal lines depend
on the longitudinal size of the system and the applied boundary
condition. The buckling strain for the supported boundary condition
is found to be independent of the longitudinal size  and estimated
to be 0.86$\%$. From a calculation of the free energy at finite
temperature we find that the equilibrium projected two-dimensional
area of the graphene nano ribbon is less than the area of a flat
sheet. At the optimum length the boundary strain for the supported
boundary condition is 0.48$\%$.

\end{abstract}
\maketitle

\section{Introduction}
Graphene, is a newly discovered almost flat one-atom-thick layer of
carbon atoms which exhibits  unique electronic properties and
unusual mechanical properties~\cite{lee,Giem2008}. Recent
experiments showed that compressed rectangular monolayer graphene on
a substrate with size $30\times100~ \mu m^2$  is buckled at about
0.7$\%$ strain~\cite{compressionamall}. Moreover tensional strain in
monolayer graphene affects the electronic properties of graphene. The
strain can generate a bulk spectral gap in the absence of
electron-electron interactions as was found within linear elasticity
theory, and a tight-binding approach~\cite{castroneto}. Different
morphological patterns of carbon nano-structures subjected to
external stress were obtained by using atomistic simulations
~\cite{yakobson}. Furthermore, atomistic simulations showed that the
Young modulus and the fracture strength decrease only weakly with
the width of the graphene nano ribbon~\cite{physlettA}.

In this paper we study the deformations and the stability  of
rectangular monolayer graphene nano ribbons  (GNR) subjected to
axial stress using atomistic simulations and the Jarzynski theorem
to calculate  the free energy~\cite{jar}. Recently, Colonna
\emph{et al} applied the free energy integration based method to
explain the melting properties of
graphite~\cite{freeenergyfasolino}. We will compare the obtained
critical buckling force with the one  predicted by elasticity
theory. We found several longitudinal deformation modes and predict
that the axial buckling boundary strain is independent of the size
in the case of laterally supported GNRs. Moreover, from a
calculation of the free energy, uncompressed GNR is
thermodynamically less stable than the GNR at the buckling
threshold. But the buckled state is less stable than the GNR at its optimum length.\\

This paper is organized as follows. In Sec.\,2  we  introduce the
atomistic model and, the simulation method. Section\,3 contains a
discussion of the elasticity theory predictions for both free
boundary condition and laterally supported boundary condition.
 We also give the buckling thresholds and the obtained deformations and compare
 our results to those from elasticity theory.
Results for the Young's modulus and pre-stresses are presented and compared to  available experimental results.
 The stability of buckled GNRs are studied in the last part of Sec.\,3.
In Sec.\,4 we will conclude the paper.

\section{Method and model}
\begin{figure*}
\begin{center}
\includegraphics[width=1.\linewidth]{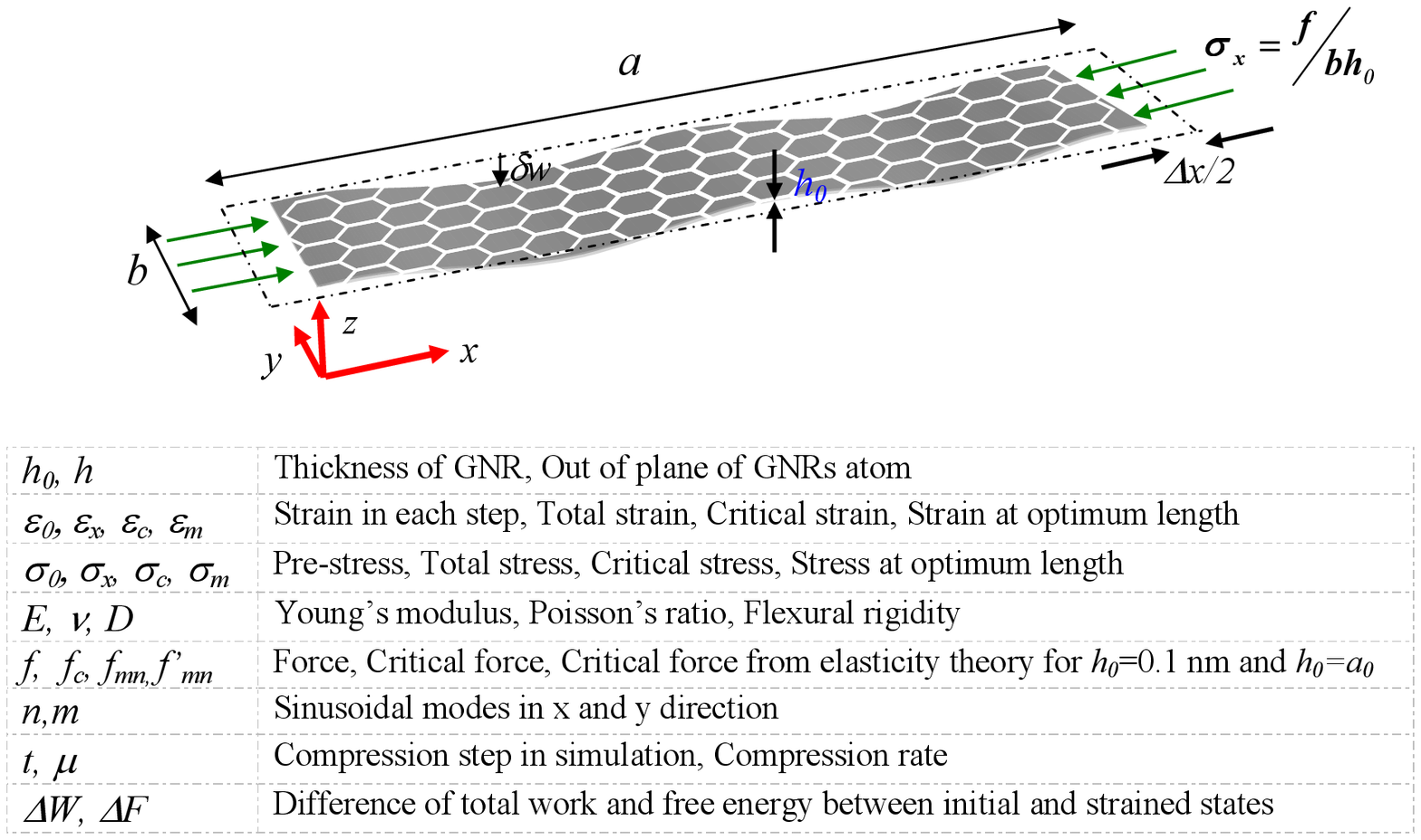}
\caption{(Color online) Schematic model for a plate under axial
strain with free boundary condition (a model for elasticity theory),
 dashed-dotted rectangle is the initial non-compressed plane. We list all relevant variables describing the GNR. \label{figmodel} }
\end{center}
\end{figure*}

 Classical atomistic molecular dynamics
simulations (MD) is employed  to simulate compressed GNRs using
Brenner's bond-order potential~\cite{brenner}. Our system is a
rectangular GNR with dimensions $a\times b$, in $x$ and $y$
directions with armchair and zig-zag edges, respectively. For
simplicity we set the lateral dimension $b=10n_y\sqrt{3}a_0$~where
$a_0$=0.142~nm, $n_y=3,4$~and the longitudinal dimension $a=30n_x
a_0$ with $n_x=2,3,...10$. In Fig.~\ref{figmodel} we depict a
schematic model for a GNR under axial strain with free boundary
condition (at $y=0$ and $y=b$) and list all relevant variables
describing GNR under axial strain. Note that $n_x$ and $n_y$ are two
integer numbers that are related to the number of atoms in the
armchair and zig-zag direction, respectively. The corresponding
values for length and width of GNRs can be found in
Table~\ref{tablea}.

\small{\begin{table*}
\begin{tabular}{|c|c|c|c|c|c|c|c|c|c|||c|c|c|}
\hline
$n_x$&2&3&4&5&6&7&8&9&10&$n_y$&3&4\\
\hline
$a(\AA)$&85.2&127.8&170.4&213&255.6&298.2&340.8&383.4&426&$b(\AA)$&98.38&122.96\\
\hline
\end{tabular}
  \centering
\caption{Length $a$ and width $b$ of GNRs which are related to the integer numbers $n_x$ and $n_y$ through $b=10n_y\sqrt{3}a_0$~ and  $a=30n_x a_0$
 where $a_0$=0.142~nm.}\label{tablea}
\end{table*}}
Initially the coordinates of all atoms are put in a flat surface of
a honey comb lattice with nearest neighbor distance equal to
$a_0$~and the initial velocities were extracted from a
Maxwell-Boltzman distribution at the given temperature. Before
starting the compression, the system is equilibrated during 50 ps
(100.000 time steps). Compressing direction is always $x$ and two
rows of atoms in both right and left edges, $x=0$ and $x=a$, are
fixed during the compression steps ($\delta x=0.04$~\AA) with the
rate $\mu=1.6$ m/s. The boundary axial strain after $t$
compression steps is
\begin{equation}
 \epsilon_x=t \epsilon_0,~~\epsilon_0=2\delta x/(30 a_0 n_x),\label{step}
\end{equation}
where $\epsilon_0=\frac{0.188}{n_x}\%$ is the strain after a single
compression step. After each compression step, we wait 2.5~ps to
allow the system to relax. For the edges at $y=0$ and $y=b$, we used
the supported boundary condition and the free boundary condition. We
simulated the system at room temperature and employed a Nos'e-Hoover
thermostat. \\
\begin{figure*}
\begin{center}
\includegraphics[width=0.3\linewidth]{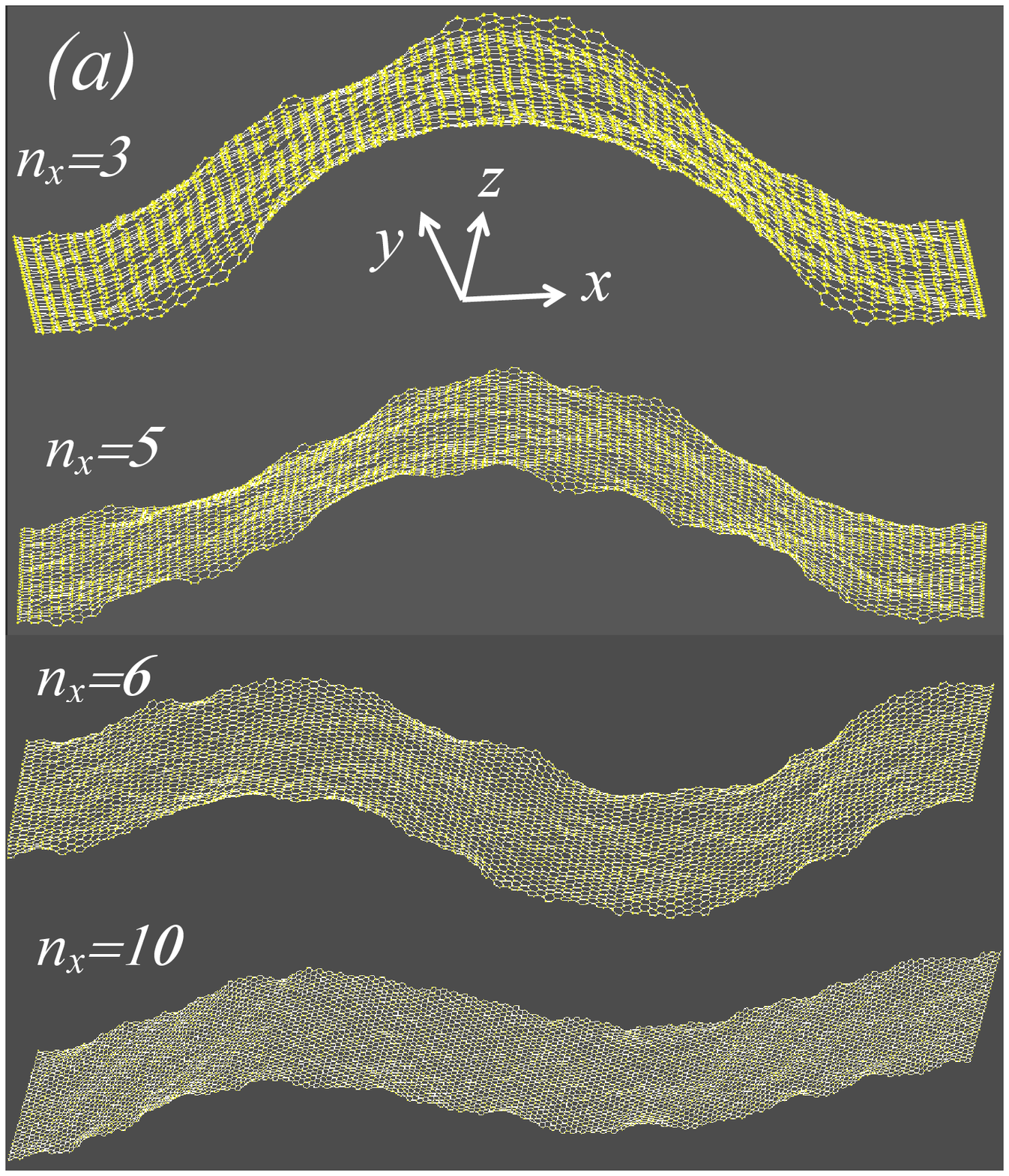}
\includegraphics[width=0.3\linewidth]{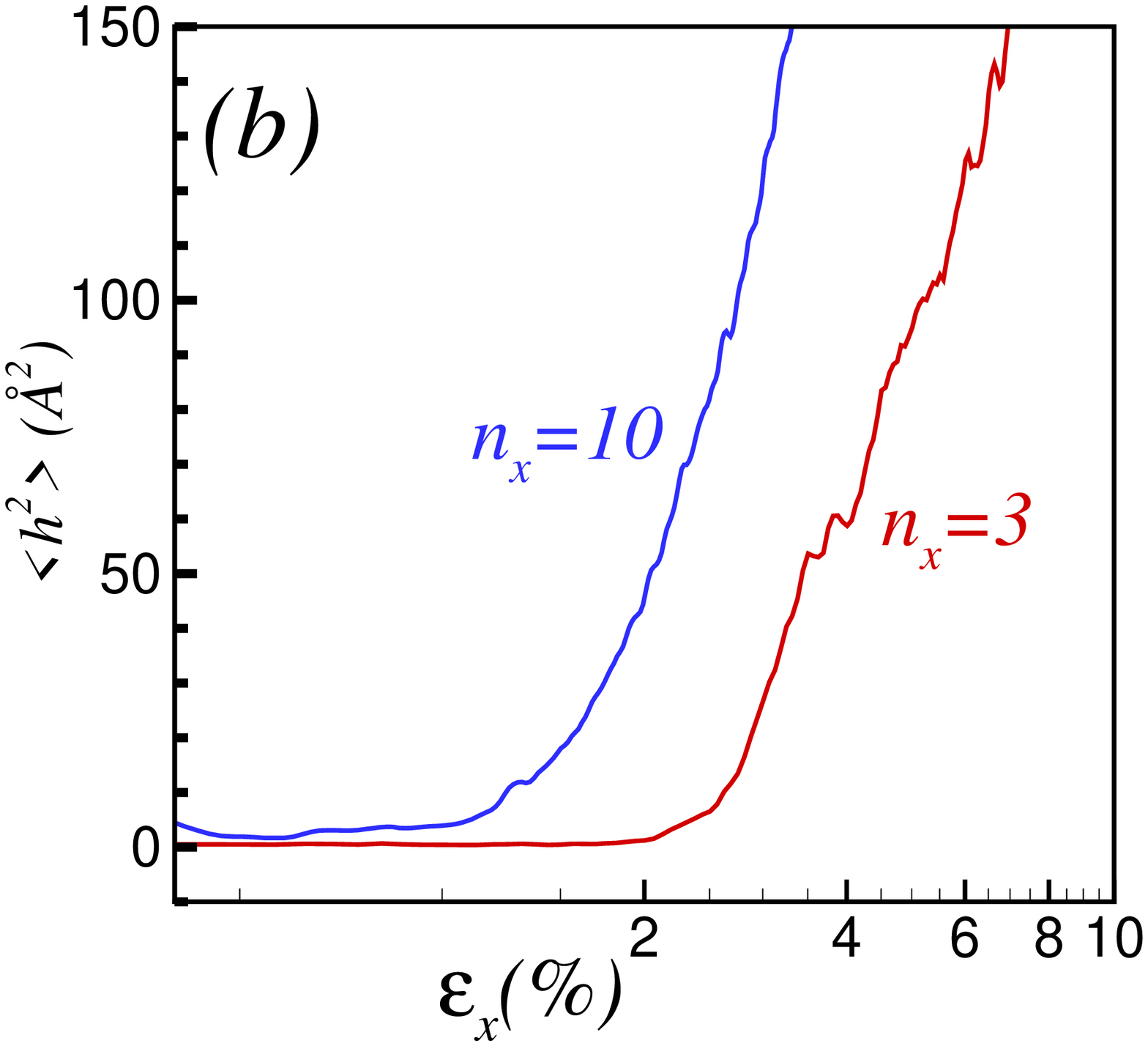}
\includegraphics[width=0.3\linewidth]{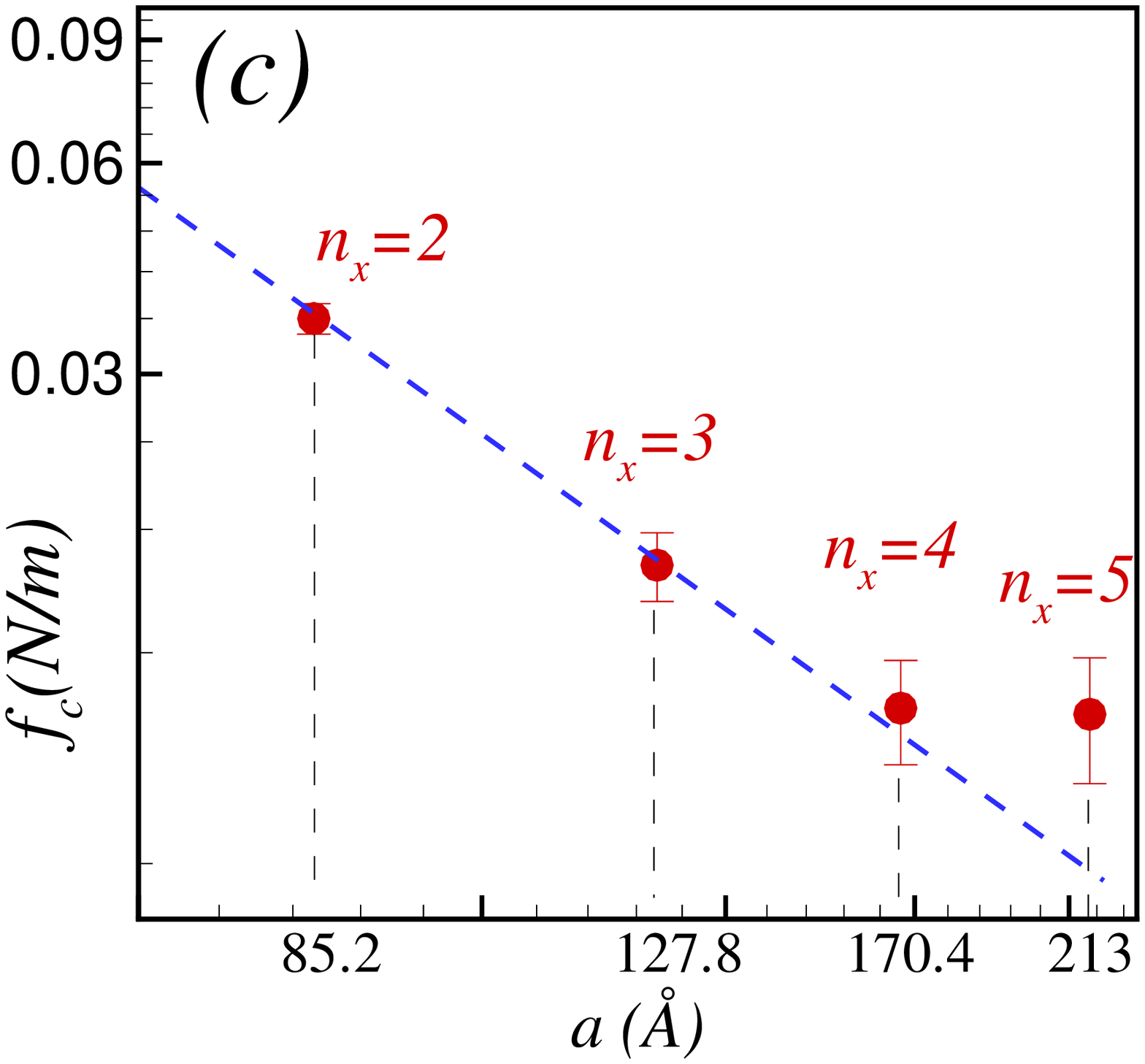}
\caption{(Color online)(a) Buckled graphene nano-ribbons for four
different length $n_x$. Here $\Delta x=$0.768 nm. (b) Variation of
the out of plane displacement in log-scale for GNRs with two
different lengths versus  strain. (c) Symbols are the calculated buckling
force in log-log scale and dashed line is the curve $f_c=\frac{\pi^2
P}{a^2}$. \label{figFreeload} }
\end{center}
\end{figure*}

\section{Buckling graphene nano-ribbons and comparison with
elasticity theory}
 \subsection{Free boundary condition} For a simple bar with length $a$, under an
axial symmetric load applied at its ends, classical Euler's column
equation describes basically the buckling problem~\cite{book}.
Governing differential equation for the deflection value, $\delta
w$, becomes the harmonic oscillator equation
\begin{equation}
\delta w''+\kappa^2\delta w=0. \label{Euler}
\end{equation}
Here $\kappa^2=f_c/P$ where $f_c$ is the buckling force (or critical
force), $P$ is a parameter which is related to the Young's modulus
and the moment of inertia of the rod cross-sectional axis that is
perpendicular to the  buckling plane.
 The solution is $\delta w=A\sin(\kappa x)+B\cos(\kappa x)$.
For the boundary condition with zero deflection at the ends, we have
$\delta w=A\sin(\frac{n \pi}{a} x)$ which are sine waves.
Substituting $\kappa=\frac{n \pi}{a}$  yields the buckling force,
\begin{equation}
f_c=\frac{n^2\pi^2 P}{a^2}~~~~(n=1,2,...). \label{fc}
\end{equation}
The buckling stress can be written as $\sigma_c=\frac{f_c}{S}$,
where $S$ is the area of the cross  section of the bar. If the bar
is thin enough or long enough buckling can happen elastically
independent of the type of material~\cite{book}.

 The shape of the lowest mode ($n$=1), after and close to the
buckling threshold, is a half sine wave. Higher modes are possible
only if the column is physically constrained from buckling into the
lower modes by supporting mid points~\cite{book}.

For GNRs under axial compression with free boundary condition  we
focused on the system with $n_y$=3. After many compression steps
GNRs starts to buckle, but the shape of the deformed GNRs depends on
their size. Figure~\ref{figFreeload}(a) shows some snap shots for
the deformed GNRs with various sizes when $\Delta x=0.768$ nm, i.e.
 beyond the buckling threshold. We will discuss
on the obtained shapes later. By measuring the buckling threshold,
we find the forces which cause a sudden change in the shape of GNRs.
This threshold is found from a direct visualization of the
nano-ribbon, and in addition from the sudden increase in the average
out of plane displacement of the GNR atoms ($\langle h^2\rangle$).
The variation of $\langle h^2\rangle$ versus $\epsilon_x$ for
systems with $n_x=3$, and 10 are shown in Fig.~\ref{figFreeload}(b).
The longer GNRs has a smaller buckling strain, $\epsilon_c\simeq
1.2\%, 2.1\%$ for $n_x=10$, 3, respectively.

On the other hand, as can be seen from the simulation snap shots in
Fig.~\ref{figFreeload}(a), even for large samples, in contrast to
the large deformations along the $x$ direction, the deformations
along the $y$ direction for each $x$ value and also deformations at
the boundaries ($y$=0 and $y=b$) for each
 $x$ value are negligible. Therefore the rod assumption for GNRs
 under axial strain is a good approximation.
  Now, considering the GNR as a rod with length $a$ and estimating the buckling forces, allow us to
calculate the variation $f_c$ versus GNR length, i.e, $a$.
Furthermore, note that throughout the present paper, we calculate
the force per width.  Since for $n_x\leq5$ we found only
deformations with $n$=1 in the beginning of the buckling, hence
using Eq. (\ref{fc}) is justified. Fig.~\ref{figFreeload}(c) shows
 the variation of the critical
load versus $a$ in log-log scale where the dots are from our
simulation
 results and the dashed line is
$\frac{\pi^2 P}{a^2}$. Here we used Eq.~(\ref{fc}) for the buckling
force, but, since our system is not a rod, the definition of an
axial moment of inertia is not meaningful. Therefore, we define an
effective bending constant for GNRs, i.e. $P$. Taking the parameter $P$
 as a fitting parameter, we
found $P\simeq0.56$ eV. The corresponding number for a covalent
carbon bond is $4.0$ eV~\cite{covalantbond1}. Clearly the bending
stiffness for a single covalent C-C bond should be much larger than
a long rod made of GNR.

Here we discuss the  shapes of the deformation. As we mentioned
above, at the start  of the buckling the shape of systems with
$n_x\leq5$, is found to be almost the same as for a buckled rod
weith $n$=1, which is similar to half sine waves. Note that we
should neglect the few rows of atoms at both ends which are fixed
during the compression resulting in flat ends (see section II). By
continuing the compression up to the time 0.5 ns, GNRs acquire a
parabolic  shape. The top two pictures in Fig.~\ref{figFreeload}(a)
show two snap shots of the obtained deformations after the buckling
threshold when $\epsilon_x=2\times\frac{0.34}{n_x}=6\%, 3.6\%$ for
$n_x=3, 5$, respectively. For a fixed reduction in the length,
$\Delta x$, the system with larger $n_x$ has a smaller amplitude,
i.e. 2.609 nm and 3.412 nm for $n_x$=5 and $n_x$=3, respectively.
For larger GNRs ($n_x\geq6$), i.e.
 $a$ is  larger than 21.3 nm, the deformations (two bottom figures in Fig.~2(a)) firstly show
the next higher mode with $n\simeq2$  and after a longer simulation
time (depending on the size) they transit  slowly to the mode $n=1$.
When the length of GNRs exceeds almost 22 nm, half of this size is
comparable to the chrematistic length of graphene, i.e. 8-10
nm~\cite{fasolinonature}. Since the chrematistic length is a measure
of the range over which deformations in one region of graphene are
correlated with those in another region, so the applied boundary
stresses on the edges do not affect the system beyond this
characteristic length and in the beginning of the buckling we expect
the $n\simeq2$ mode.
 For GNR with $n_x\geq6$ we did not find a simple relation between the critical buckling force and
the length. For larger systems ($n_x>10$) the deformations are no
longer sine waves at least during our simulations time.

\begin{figure}
\begin{center}
\includegraphics[width=0.8\linewidth]{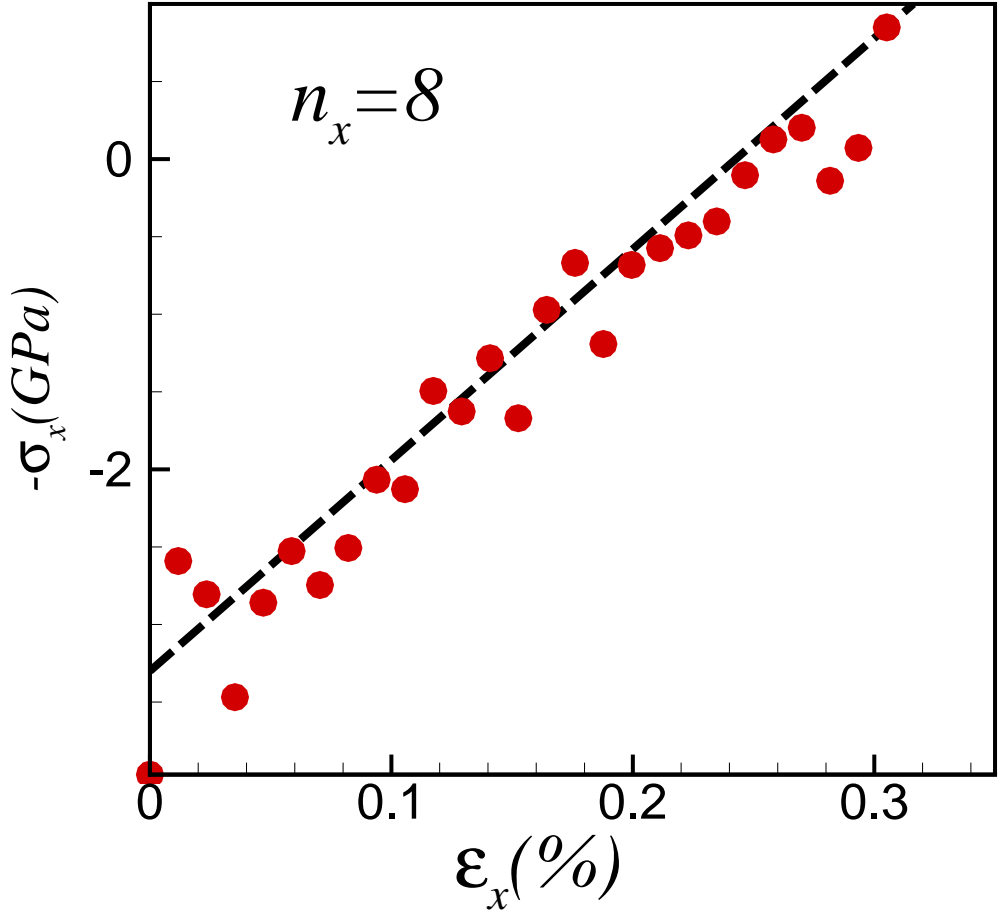}
\caption{(Color online) Stress-strain curve for the GNR with $n_x$=8
under axial stress before the buckling threshold where.
\label{stress} }
\end{center}
\end{figure}

Before ending this section, we calculate the stress-strain curve.
Before the buckling threshold, we found a linear relation between
stress and strain, i.e. $\sigma_x=E\epsilon_x+\sigma_0$, where
$\sigma_0$ is the pre-stress in the system~\cite{lee}. The linear
relation is valid for  small strains~\cite{lee,poisson}. In our
simulations,  when the GNRs are not flat and thus not compressed,
they are not in equilibrium and some boundary tension exist, i.e.
pre-stress~\cite{lee}. For instance, when $n_x=8$ we calculate the
applied stress on the right hand side (RHS) edge using
$\frac{f}{bh_0}$  (by using the thickness of graphene equal to
$h_0$=0.1 nm~\cite{Saitoh}) and we show the obtained stress-strain
curve in Fig.~\ref{stress}. The dashed line is the fitted line. The
slope of this line gives us Young's modulus 1.3$\pm$0.07 TPa and
$\sigma_0$=-3.3$\pm$0.2 GPa (negative sign indicates the direction
of compression, i.e. -\emph{x}). For the other GNRs we found Young's
modulus in the same range (e.g. $E=1.1\pm$0.08 TPa and
$\sigma_0$=-3.3$\pm$0.2 GPa for $n_x=10$
etc). These numbers are comparable to those found experimentally~\cite{lee}.\\

\subsection{Supported boundary condition} For a rectangular
plate subjected to the supported boundary condition (when movements
at $x$=0, $x=a$,  $y=$0 and $y=b$ along $y$ and $z$ directions are not allowed),
elasticity theory~\cite{book} tells us that the governing equation
for the buckled rectangular under uniform compressive axial load
per width ($f$) in the $x$-direction, can be written as
\begin{equation}
D\nabla^4 (\delta w)+f(\delta w_{xx})=0, \label{BEQ}
\end{equation}
where  $\delta w$ is the transverse deflection, $\delta w_{xx}$ is
the corresponding curvature and $D=Eh_0^3/(12(1-\nu^2))$ is the
flexural rigidity of the plate with thickness $h_0$ and Young's
modulus $E$. The general solution for the deflection can be written
as a double Fourier series
\begin{equation}
\delta w=\sum^{\infty}_{m,n=1}\tilde{w}_{mn}\sin(n\pi x/a)\sin(m\pi
y/b), \label{dw}
\end{equation}
where $(m,n)$ are integers in order to satisfy the supported
boundary condition and $\tilde{w}_{mn}$ is the amplitude of each
mode $(m,n)$. Including the appropriate strain energy and using
Eq.~(\ref{dw}), buckling occurs when~{\cite{book}
\begin{equation}
f_{mn}=\frac{\pi^2 a^2 D}{n^2}[(\frac{n}{a})^2+(\frac{m}{b})^2]^2
.\label{k}
\end{equation}
Lowest value of $f_{mn}$ with respect to the two discrete variables
$m,n$ gives the buckling force, $f_c$. The minimum buckling force
 for the considered systems always occurs for $m=1$
  and various values of $n$. It is equivalent
to a single half wave in the lateral $y$-direction and various
harmonics $n$ in the compression direction, i.e., $x$.

\begin{figure*}
\begin{center}
\includegraphics[width=0.3\linewidth]{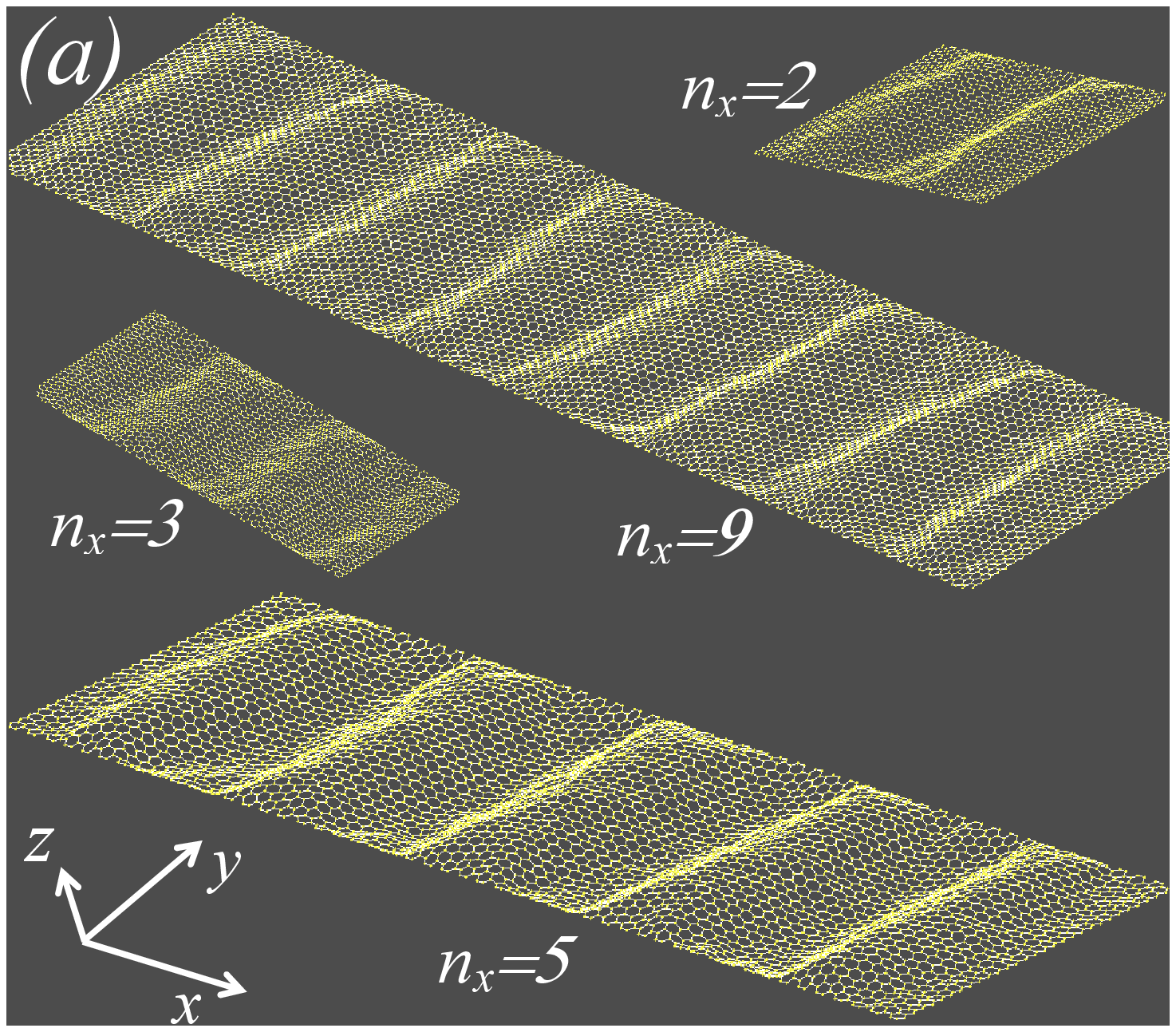}
\includegraphics[width=0.3\linewidth]{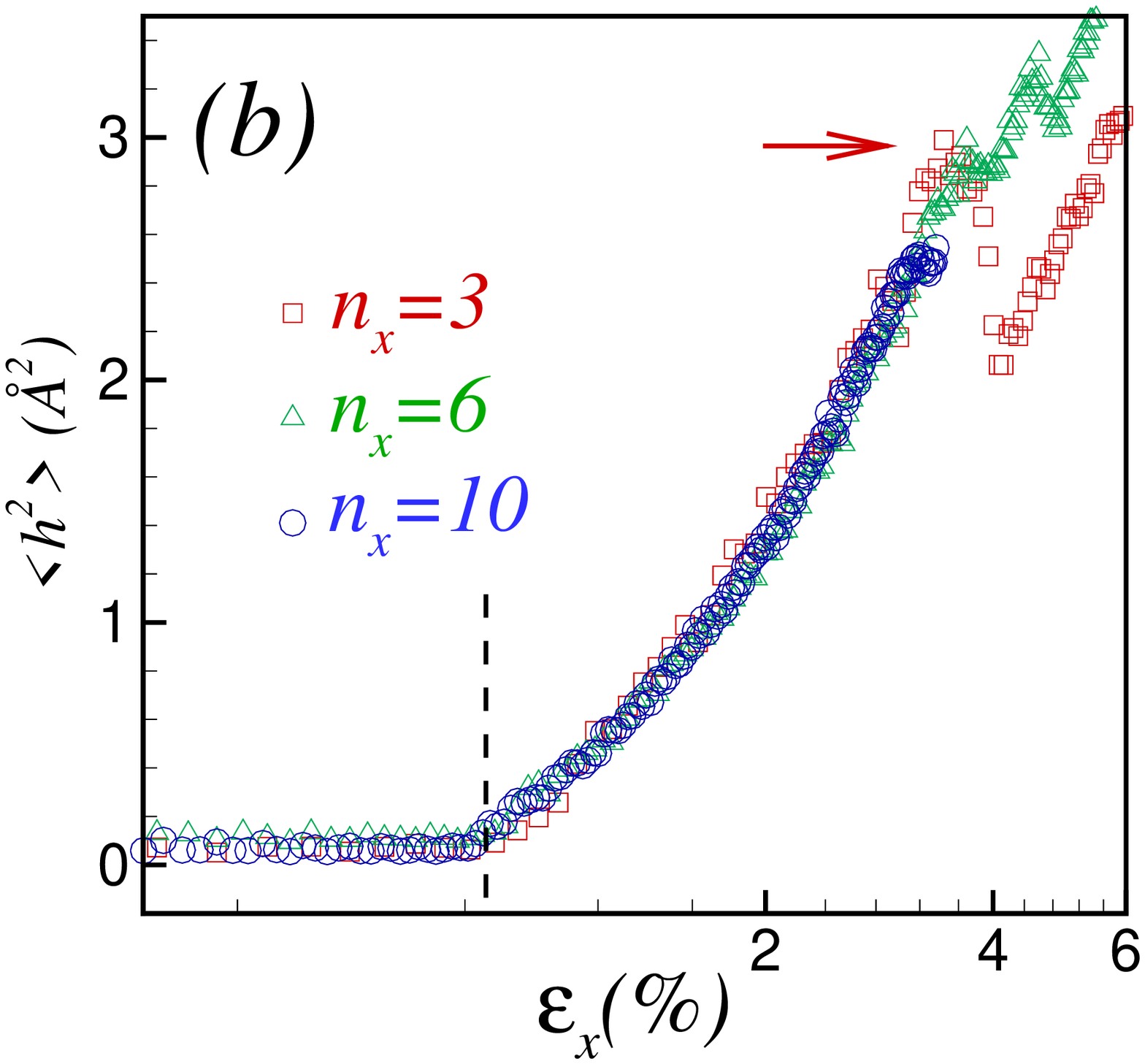}
\includegraphics[width=0.3\linewidth]{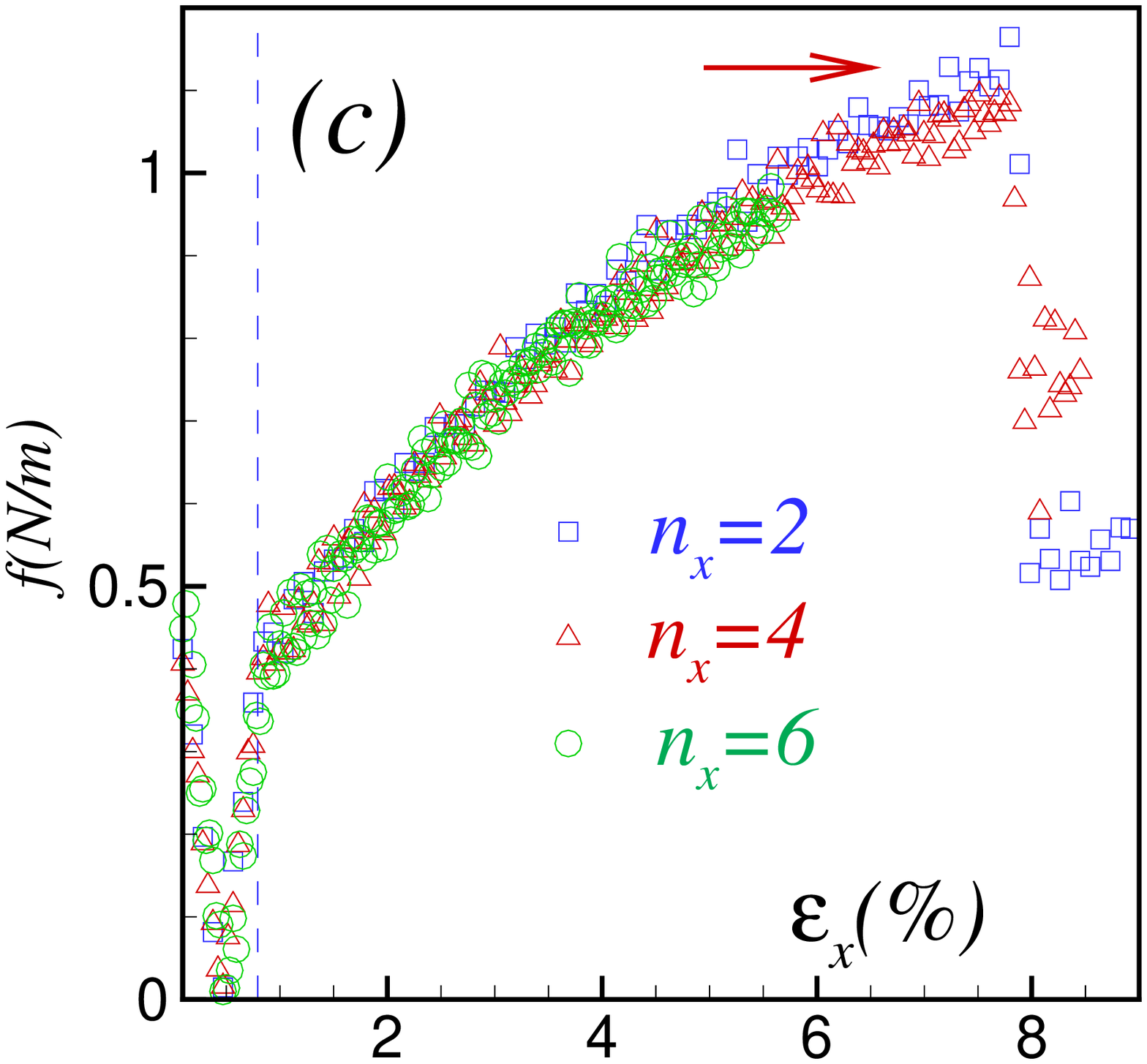}
\caption{(Color online) (a) Buckled graphene nano-ribbons for four
different length subjected to a lateral supported boundary
condition. Here applied strains are larger than the buckling
strains. (b) Variation of the out of plane displacement in log-log
scale for GNRs for three different lengths versus strain. (c) Force
applied on the boundaries for three different sizes: $n_x=2, 4, 6$.
Horizontal arrow show the instability point and vertical dashed line
shows the buckling point.\label{figsupmode} }
\end{center}
\end{figure*}

We performed several atomistic simulations for GNRs under supported
boundary condition for different $n_x$ and fixed $n_y$=3. In
Fig.~\ref{figsupmode}(a), we depict four typical snap shots for
different sizes of the buckled GNRs (where $\epsilon_x$ is always
larger than the buckling strain, $\epsilon_c$). As can be seen from
this figure, those satisfy the condition $m$=1, and by increasing
the size of the system, we obtained higher longitudinal modes.
Furthermore, higher strains ($\epsilon_x>\epsilon_c$) increase the
amplitude of the deformations and increase the number $n$ slightly
(usually $n'=n+1,n+2$). Similar to the free boundary condition case,
the buckling thresholds, for smaller $n_x$, are obtained at smaller
axial strains. In Table \ref{table1}, we list our calculated
buckling forces per width ($f_c$), buckling strains and the
prediction from elasticity theory according to Eq.~(\ref{k}). To
calculate $f_{mn}$, we used $E=340$ N/m$^2$ and
$\nu=0.165$~\cite{lee} with $h_0\sim0.1$~nm, as a typical thickness
for GNR. Larger thickness yields a larger $f_{mn}$ so that $h_0=a_0$
yields a better agreement between $f'_c$ and $f_{mn}$.  Note that
elasticity theory is not  applicable to GNRs under
strain~\cite{compressionamall}  for the critical values for force or
strains. For instance, although elasticity theory predicts that the
very small thickness of graphene yields a zero flexural rigidity but
the bond-angle effects on the interatomic interactions of graphene
(three body terms in Brenner's potential) gives a non zero flexural
rigidity for graphene~\cite{compressionamall}  (see Refs.~[28,29] in
Ref.~\cite{compressionamall}). Therefore, because of the non-zero
flexural rigidity in graphene, larger critical forces with respect
to the predictions from elasticity theory, are to be expected.

 Figure~\ref{figsupmode}(b) shows
the variation of $\langle h^2\rangle$ versus $\epsilon_x$ for
systems with $n_x=3, 6, 10$ when $n_y=3$. Vertical dashed line shows
the transition points to the buckled state. Clearly the behavior of
$\langle h^2\rangle$ is the same for all cases. As we see,
surprisingly, the buckling strain is independent of the longitudinal
size (vertical dashed line in Fig.~\ref{figsupmode}(b)) of GNRs and
it varies in the range [0.84-0.89]$\%$ while for larger width, e.g.
$n_y=4$, the buckling strain varies in the range [0.8-0.93]$\%$. The
average for $n_y=3$ is 0.8688$\%$ and for $n_y=4$ is 0.8677$\%$
which are very close. Therefore we conclude that the buckling strain
is independent of the longitudinal dimension and  depends only
weakly  on the width of the GNR.

It is important to note that higher $\epsilon_x$
 (especially for the smaller system with small $n_x$) results in instabilities
 which makes the GNRs partially crumple (see Figure~\ref{figinstability}).
 This is due to the sp$^{2}$ bond breaking at the nonuniform deformed (crumpled) region of
 edges. Red horizontal arrow in Fig.~\ref{figsupmode}(b) indicates the instability point for $n_x=3$.
  To find  the instabilities and the buckling thresholds, we calculated the variation of the boundary forces
versus $\epsilon_x$. The force on the left hand side (LHS) edge is
shown in Fig.~\ref{figsupmode}(c). The forces on the LHS edge
decrease to zero before the buckling thresholds which mean that GNRs
at those points have an optimum length where the system does not
feel any external forces (Fig.~\ref{figinstability}(a)). We will
return to this point later. In Fig.~\ref{figsupmode}(c) the vertical
dashed line indicates the buckling threshold and the arrow gives the
first sudden changes in the force (for three systems with $n_x=2, 4,
6$) which indicates an instability in the shape of the GNR.

As we mentioned above, Fig.~\ref{figinstability} shows three snap
shots of a GNR (having $n_x=2$ size) at three different strains. For
Fig.~\ref{figinstability}(a) strain is $\epsilon=\epsilon_m$ (strain
at optimum length). Two other snap shots are  GNR before
(Fig.~\ref{figinstability}(b)) and after
(Fig.~\ref{figinstability}(c)) the first instability point in
Figs.~\ref{figsupmode}(b,c).
 The strain (and compression steps) in Fig. \ref{figinstability}(c) is larger than in Fig.~\ref{figinstability}(b). Both
strains in Figs. \ref{figinstability}(b,c) are larger than the
critical strain ($\epsilon_c$). Brenner's potential is not
responsible for the occurred non-hexagonal structures (bond breaking
effects) in some nonuniform deformed regions of
Fig.~\ref{figinstability}(c). The reason is that in common covalent
potentials, people use a drastic reduction in cut-off distance (here
$\simeq$2\AA) which has resulted in  a good description of the
material before fracture or bond breaking. But, it leads to an
overestimation of critical loads and shear stresses in fracture
mechanics and tribology, where bond breaking occur~\cite{Mardar}.
 Here we do not study the bond breaking situation
or fracture mechanism that can occur in  GNRs by a continues
compression beyond the buckling state. Modification to Brenner's
potential have been proposed in order to include fracture mechanisms
and bond breaking situations~~\cite{modificationBRENNER}. The idea
to those modifications is to find nearest neighbors by a criterion
other than distance~\cite{modificationBRENNER} and this  by
employing empirical screening functions as introduced in ref ~
~\cite{modify}.

\begin{figure}
\begin{center}
\includegraphics[width=0.8\linewidth]{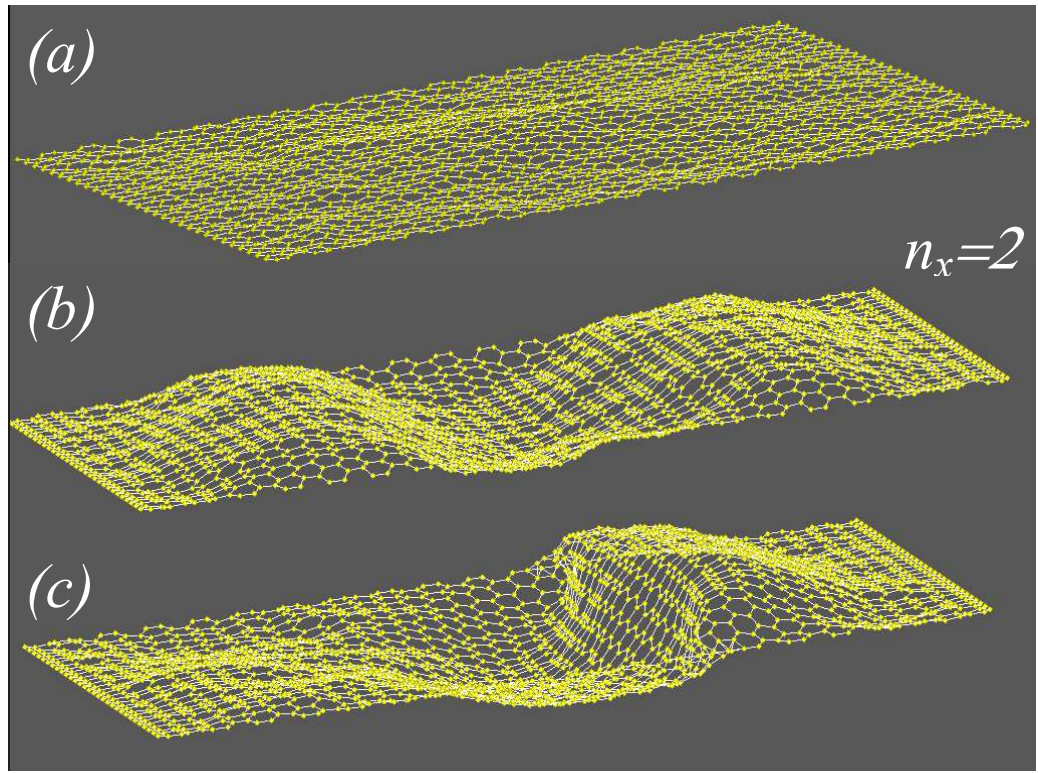}
\caption{(Color online) GNR with $n_x=2$ size at its optimum length
(a), and before (b) and after (c) the
instability.\label{figinstability} }
\end{center}
\end{figure}
\begin{table}
\begin{tabular}{|c|c|c|c|c|c|c|c|c|c|c|}
    \hline
 $n_y$&$n_x\rightarrow$&2&3&4&5&6&7&8&9&10\\
  \hline
  \hline
&$n$&3&4&5&7&9&11&11&14&14\\
&$f_c$&0.45&0.43&0.47&0.47&0.40&0.43&0.37&0.4&0.36\\
$3$&$f_{mn}$&0.14&0.12&0.11&0.13&0.14&0.15&0.12&0.15&0.13\\
&$f'_{mn}$&0.40&0.34&0.31&0.36&0.40&0.42&0.35&0.42&0.36\\
&$\epsilon_c(\%)$&0.84&0.87&0.89&0.89&0.88&0.84&0.85&0.88&0.88\\
  \hline
 &    $n$&2&3&5&6&7&8&9&11&12\\
&$f_c$&0.38&0.37&0.34&0.33&0.32&0.34&0.32&0.32&0.32\\
$4$&$f_{mn}$&0.07&0.07&0.09&0.09&0.08&0.08&0.08&0.09&0.09\\
&$f'_{mn}$&0.19&0.19&0.24&0.25&0.24&0.23&0.22&0.25&0.25\\
&$\epsilon_c(\%)$&0.90&0.93&0.93&0.91&0.87&0.89&0.87&0.8&0.80\\
 \hline
\end{tabular}
  \centering
\caption{The periodicity number ($n$) of the sine waves observed in
the buckled laterally supported GNRs in the longitudinal direction
versus the length of the GNR. The calculated buckling force from our
MD calculations ($f_c$) and results from elasticity theory according
to Eq.~(\ref{k}), $f_{mn}$, for two widths $n_y=3, 4$. When
calculating $f_{mn}$ we used $h_0=0.1$ nm and $h_0=a_0$ was used for
$f'_{mn}$. $\epsilon_c$ is the strain at which the GNR
buckles.}\label{table1}
\end{table}

\subsubsection{\textbf{Static buckling deformations and the stability of
buckled GNRs}}

 In the last part of this paper, we calculate the
change of the free energy of laterally supported GNRs subjected to
axial strain. Due to the application of an external force on the
boundary, an equilibrium approach is no longer applicable and a
non-equilibrium MD is needed. Independent of the path and for a
finite evolution rate, Jarzynski found an equality between the
difference of the free energy and the total work done on the system
($W$) during a non-equilibrium evolution~\cite{jar}
\begin{equation}
\Delta F=-\beta^{-1}\ln\langle{\exp(-\beta W)}\rangle,\label{jeq}
\end{equation}
where $\beta=1/k_BT$. The averaging is done over the realization of
the switching process between the initial and the final state.
Equation (\ref{jeq}) makes a connection between the difference of
the equilibrium free energy and the non-equilibrium work.

Using Eq.~(\ref{jeq}) we calculated the changes  of the free energy
when compressing the GNR of size $n_x=10$ and plot the results in
Fig.~\ref{freeenergy}(a). The inset shows the
 total work done on the system for 10 simulations with
different initial conditions. Comparing the curve for $\langle W
\rangle$ and $\Delta F$ shows that the difference of the free
energies are smaller than the total work which agree with
 $\langle W \rangle \geq \Delta F$ as can be derived from Eq.~(\ref{jeq})~\cite{jar}.
Fig.~\ref{freeenergy}(b) shows the change in $W$ for systems having
different values for $n_x$. The minimum in the free energy curve
corresponds to an equilibrium length (also to an amount of strain
($\epsilon_m$) where there is zero force on the boundaries at $x=0,
x=a$ (see Fig.~\ref{figsupmode}(b)).

Notice that our non-compressed GNRs (in the beginning of the
simulations) are flat honeycomb lattice structures which are not  in
a thermo-mechanically equilibrium state at finite temperature.
Therefore the free energy of this state should be higher than the
equilibrium state. It is well known that at finite temperature the
equilibrium state of suspended graphene is not exactly a flat sheet
and some intrinsic ripples are present~\cite{fasolinonature}. At the
optimum length our suspended  GNRs are rippled
(Fig.~\ref{figinstability}(a)) and the system is in the equilibrium
state. Figure~\ref{freeenergy}(a) shows the variation of the total
free energy versus applied strain.
 As we see from Fig.~\ref{freeenergy}(a)
the rippled state (minimum point in free energy curve) has a lower
free energy with respect to the initial non-compressed GNRs,
($\Delta F=$-2.27 eV). The inset of Fig.~\ref{freeenergy}(a) depicts
the corresponding change in the total performed work on the system
for 10  simulations with different initial conditions. On the other
hand, since the optimum length of the suspended GNR is less than the
initial non-compressed length ($a$), we may conclude that at finite
temperature the projected 2D-area ($a(1-\epsilon_m)\times b$) of the
GNR is less than the area of a flat GNR ($a\times b$). Surprisingly,
as we see from Fig.~\ref{freeenergy}(b) the boundary strain at the
optimum length $\epsilon_m=0.48\%$ is size independent. The vertical
dashed lines in Figs.~\ref{freeenergy}(a,b) indicate the buckling
strain, i.e. $\epsilon_c$ and the strain at the optimum length, i.e.
$\epsilon_m$ respectively. Free energy in the buckling point is less
than the free energy of the initial non-compressed system but it is
higher
 than the free energy  for the rippled state (minimum value).
Difference between the free energy  of the rippled state and the
buckled state is -1.87 eV.
 For the $n_x=10$ system we stopped the
compression after the buckling threshold and
 equilibrated the system during a  very long time and found static and stable
sine wave deformations in the GNR.

\begin{figure}
\begin{center}
\includegraphics[width=0.8\linewidth]{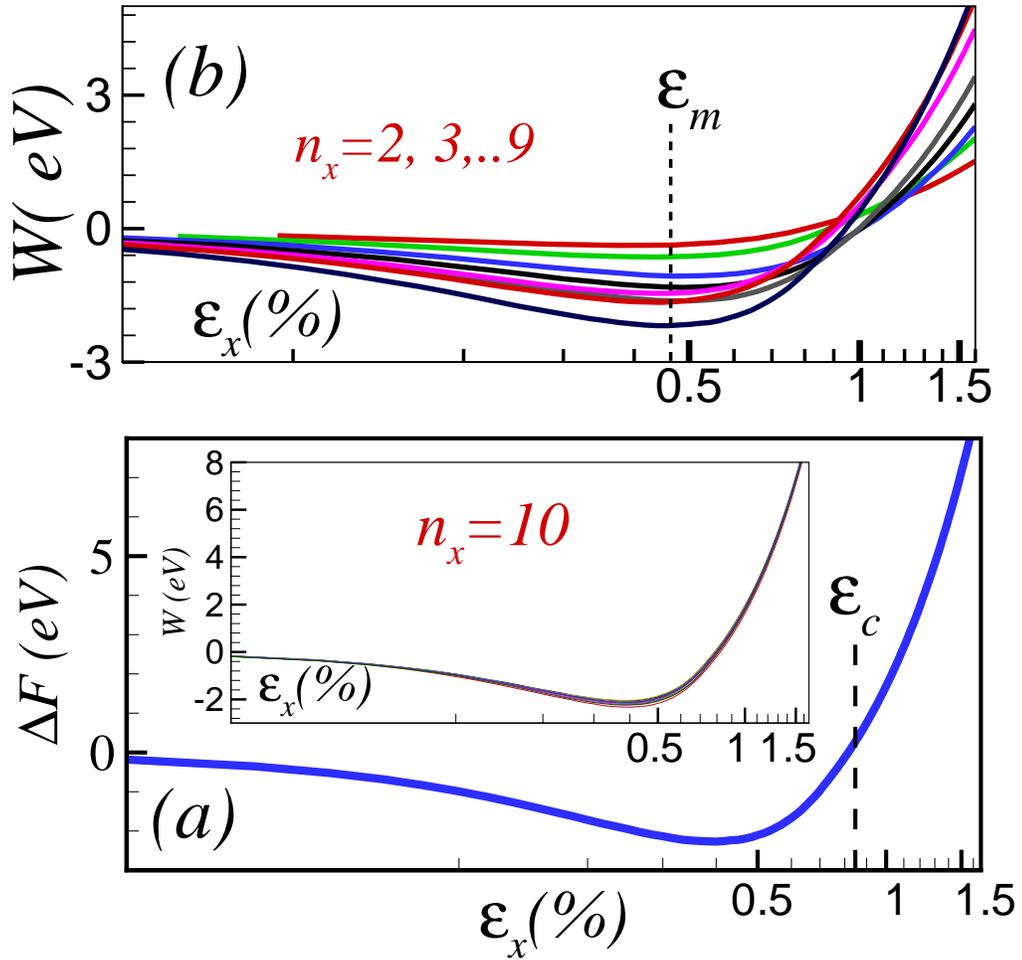}
\caption{(Color online) (a) Change of the free energy during
compression for a system with size $n_x=10$ versus  strain. The
inset depicts the corresponding change in the total performed work
on the system for 10  simulations with different initial conditions.
 (b) Total work
done on the system for various sizes versus  strain.\label{freeenergy} }
\end{center}
\end{figure}

\section{Conclusions}
 Deformations in the graphene nano-ribbons subjected to axial boundary compression
   are static sine waves with different number of nodal lines, depending on the length of the GNRs.
The deformations predicted from elasticity theory  for the buckled
rod and rectangular plate are similar to those obtained for
 the buckled GNRs in the case of free boundary
condition and also for the laterally supported boundary condition.
However, the critical force and flexural rigidity of GNRs are larger
than predicted from elasticity theory. We found a linear relation
for the stress-strain curve for small strains (i.e. before the
buckling threshold). The buckling strain (0.86$\%$) and the strain
caused by the equilibrium length ($0.48\%$) are independent of the
longitudinal size of the system and they  depend weakly on the width
of GNRs. From the free energy of the GNRs at the buckling threshold,
we found that they are thermodynamically more stable than those
before compression, i.e. flat GNRs.\\

\emph{{\textbf{Acknowledgment}}}.
 This work was supported by the Flemish science foundation (FWO-Vl) and the Belgium Science Policy~(IAP).

\end{document}